\documentclass[aps,twocolumn,floatfix,amsmath,amssymb,showpacs, tightenlines]{revtex4}
\usepackage{amssymb}
\usepackage{graphicx}
\usepackage{epsfig,graphicx,times}
\usepackage{amstext}
\usepackage{amsmath}
\usepackage{graphicx}
\usepackage{latexsym}
\usepackage{bm}

\begin{document}

\title{Scalable superconducting charge qubit quantum computer via gate voltage and external flux}
\author{Wen Yi Huo$^{1}$ and Gui Lu Long$^{1,2,3}$ }

\address{
$^1$Key Laboratory for Atomic and Molecular NanoSciences and Department of Physics,
Tsinghua University, Beijing 100084, China\\
$^2$  Tsinghua National Laboratory For Information Science and Technology, Beijing
100084, China\\
$^3$Institute of Microelectronics, Tsinghua University, Beijing 100084, China}

\date{\today}

\begin{abstract}
We present a scalable scheme for superconducting charge qubits with
the assistance of one-dimensional superconducting transmission line
resonator (STLR) playing the role of data bus. The coupling between
qubit and data bus may be turned on and off by just controlling the
gate voltage and externally applied flux of superconducting charge
qubit. In our proposal, the entanglement between arbitrary two
qubits and $W$ states of three qubits can be generated quickly and
easily.
\end{abstract}

\pacs{03.67.Lx,85.25.-j} \maketitle

\section{introduction}
As a candidate of  scalable solid-state quantum computer, superconducting circuits with
Josephson junction  \cite{makhlin} have attracted much attention in recent years. A
series of successful experiments in single superconducting charge  \cite{nakamura},
flux  \cite{mooij} and phase  \cite{yu} qubits have demonstrated the macroscopic
quantum coherence and the relative long coherence time. Due to the good performance of
single superconducting qubit, people are now exploring the possibility for scaling up
to many qubits.

Two-qubit experiments have been performed in superconducting charge
 \cite{pashkin, yamamoto}, flux  \cite{izmalkov, majer, plourde} and
phase  \cite{xu, berkley, mcdermott} qubits and the entanglement has been observed.
Usually, the interaction between qubits in these experiments is always on during the
manipulation.  There are many theoretical proposals  \cite{makhlin, shnirman, you,
wang1, blais1, averin, cosmelli} for coupling any pair of qubits selectively through a
common data bus or by adding additional sub-circuits. There are currently experimental
efforts in  implementing these proposals.

Recently, superconducting qubits coupled to an $LC$ oscillator or one-dimensional
superconducting transmission line resonator (STLR) \cite{blais2} has attracted much
interest.  Such systems were of great interest not only in the study of the fundamental
quantum mechanics of open systems but also as the potential candidates of scalable
superconducting quantum computers. In the ion-trap quantum computer \cite{cirac}, the
harmonic oscillator played an important role as data bus. Similarly, the $LC$
oscillator or STLR can also play the role of data bus. Data-bus plays an  important
role  in quantum information processing \cite{spiller}. With the help of data bus,
state transfer \cite{wang2} and $n$-qubit controlled phase gate \cite{yang} for
superconducting phase qubits were proposed. By coupling a charge qubit with a
micro-cavity, macroscopic superposition states can be  generated by manipulating the
gate voltage and external flux in an elegant way in Refs. \cite{nori}. By using the
appropriate time-dependent electromagnetic fields, Liu {\it et al.}  \cite{liu}
presented a scheme of scalable circuit for superconducting flux qubits, where quantum
two-gate is realized by applying an external classical light fields.

The work of \cite{liu} provides a scalable superconducting quantum computer scheme
using flux qubit. Inspired by the interesting idea in Ref.\cite{liu}, we present a
scalable superconducting charge quantum computer scheme up to many qubits with the
assistance of STLR playing the role of data bus. In this scheme, both single- and two-
qubit gates can be implemented by just controlling the gate voltage and externally
applied flux. It is found that Bell states and $W$ states can be generated quickly in a
simple way.

\section{The model Hamiltonian}

\subsection{Single superconducting charge qubit}

\begin{figure}[htbp]

\includegraphics[width=1.5in]{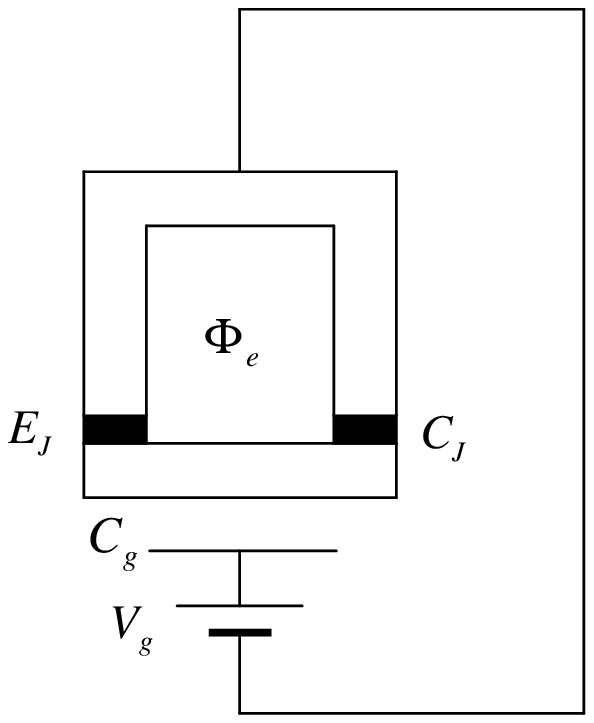}
\caption{The structure of superconducting charge qubit. The Josephson coupling energy
and capacitance of the two identical Josephson junctions is $E_J$ and $C_J$. The gate
voltage $V_g$ and externally applied flux $\Phi_e$ are used to bias the superconducting
island and the dc SQUID.}\label{qubit}
\end{figure}

The single superconducting charge qubit consists of a small
superconducting island with cooper-pair charge $Q=2ne$ connected by
two identical Josephson junctions(each with capacitance $C_J$ and
Josephson coupling energy $E_J$) to a superconducting electrode.
This is the structure of a dc SQUID. A gate voltage $V_g$ is coupled
to the superconducting island through gate capacitance $C_g$. The
gate voltage $V_g$ and externally applied flux $\Phi_e$ are
externally controlled and used to bias the superconducting island
and the dc SQUID. A schematic diagram of the single-qubit structure
is show in Fig.\ref{qubit}. The Hamiltonian of the system is
\begin{equation}
H_q=E_c(n-n_g)^2-2E_J\cos(\frac{\pi\Phi_e}{\Phi_0})\cos\varphi,
\end{equation}
where
\begin{equation}
E_c=\frac{(2e)^2}{2(2C_J+C_g)}
\end{equation}
is the single-cooper-pair charging energy of the island,
$n_g=C_gV_g/2e$ is the gate charge induced by the gate voltage,
$-2E_J\cos(\pi\Phi_e/\Phi_0)$ is the effective Josephson coupling
energy and $\Phi_0=h/2e$ is the flux quantum.

In the charge regime, that is $E_c\gg E_J$, and in the neighborhood
of $n_g=1/2$, only two charge states, say $|0\rangle$ and
$|1\rangle$, are relevant. In the representation of this two charge
states, the reduced two-state Hamiltonian can be written in the
spin-$1/2$ form
\begin{equation}\label{qh}
H_q=-\frac{1}{2}E_c(1-2n_g)\sigma_z-E_J\cos(\frac{\pi\Phi_e}{\Phi_0})\sigma_x.
\end{equation}
The eigenvalues of single-qubit Hamiltonian are
\begin{eqnarray}
E_{\pm}=\pm
E=\pm\sqrt{\frac{1}{4}E_c^2(1-2n_g)^2+E_J^2\cos^2(\frac{\pi\Phi_e}{\Phi_0})},\nonumber
\end{eqnarray}
and eigenstates
\begin{eqnarray}
|g\rangle=\sin\frac{\eta}{2}|1\rangle+\cos\frac{\eta}{2}|0\rangle,\\
|e\rangle=\cos\frac{\eta}{2}|1\rangle-\sin\frac{\eta}{2}|0\rangle,
\nonumber
\end{eqnarray}
where
\begin{eqnarray}
\eta=\tan^{-1}(\frac{2E_J\cos(\frac{\pi\Phi_e}{\Phi_0})}{{E_c(1-2n_g)}}).\nonumber
\end{eqnarray}
In the representation of eigenbasis, the Hamiltonian is diagonalized
as
\begin{equation}
H_q=-\frac{1}{2}\hbar\Omega\rho_z,
\end{equation}
where $\Omega=2E/\hbar$ and $\rho_z=|g\rangle\langle
g|-|e\rangle\langle e|$.

\subsection{Interaction of the charge qubit with one-dimensional cavity}

Here we consider the one-dimensional cavity realized by STLR
 \cite{blais2}. For an ideal one-dimensional STLR with length $l$ and
the boundary conditions $I(0,t)=I(l,t)=0$, the quantized current is
\begin{eqnarray}
I=\sum_k\sqrt\frac{\hbar\omega_k}{lL}(a_k^{\dagger}+a_k)\sin\frac{k\pi
x}{l},\nonumber
\end{eqnarray}
where $\omega_k=k\pi/(l\sqrt{LC})$ and $L$ $(C)$ is the inductance
(capacitance) per unit length. The Hamiltonian of the STLR reads
\begin{equation}
H_c=\sum_k\hbar\omega_k(a_ka_k^{\dagger}+\frac{1}{2}).
\end{equation}
Consider the $k$-th mode of the STLR, the current
\begin{eqnarray}
I_k=\sqrt\frac{\hbar\omega_k}{lL}(a_k^{\dagger}+a_k)\sin\frac{k\pi
x}{l}.\nonumber
\end{eqnarray}
Placing the superconducting qubits at the points $x=(2n+1)l/2k$,
where $n$ is an arbitrary integer, the Hamiltonian of the combined
system reads
\begin{eqnarray}\label{h1}\nonumber
H_t&=&\hbar\omega(a^{\dagger}a+\frac{1}{2})-\frac{1}{2}E_c(1-2n_g)\sigma_z\\
&&-E_J\cos(\frac{\pi(\Phi_e+\Phi_q)}{\Phi_0})\sigma_x,
\end{eqnarray}
where $\Phi_q=\mu_0S\sqrt{\hbar\omega/lL}(a^{\dagger}+a)/(2\pi d)$
is the quantized magnetic flux induced by the quantized current, $d$
is the distance between the qubit and STLR and $S$ is the area
enclosed by the dc SQUID. Here, for simplicity, we denote $\omega_k$
as $\omega$ and $a_k$ as $a$. Generally speaking, $\Phi_q\ll\Phi_e$,
then the Hamiltonian(\ref{h1}) approximately reads
\begin{eqnarray}\nonumber
H_t&=&\hbar\omega(a^{\dagger}a+\frac{1}{2})-\frac{1}{2}E_c(1-2n_g)\sigma_z-E_J\cos(\frac{\pi\Phi_e}{\Phi_0})\sigma_x\\
&&+\frac{\mu_0E_JS}{2\Phi_0
d}\sqrt{\frac{\hbar\omega}{lL}}\sin(\frac{\pi\Phi_e}{\Phi_0})(a^{\dagger}+a)\sigma_x.\nonumber
\end{eqnarray}
In the eigenbasis of the qubit's Hamiltonian and under the
rotating-wave approximation, the total Hamiltonian of the combined
system (qubit and STLR) reads
\begin{eqnarray}\label{jc}
H=\hbar\omega(a^{\dagger}a+\frac{1}{2})-\frac{1}{2}\hbar\Omega\rho_z+\hbar
\lambda(a^{\dagger}\rho_-+\rho_+a),
\end{eqnarray}
where $\rho_+=|e\rangle\langle g|$, $\rho_-=|g\rangle\langle e|$ and
the coupling coefficient of qubit and STLR
\begin{eqnarray}
\lambda=\frac{\mu_0E_JS}{2\Phi_0
d}\sqrt{\frac{\hbar\omega}{lL}}\sin(\frac{\pi\Phi_e}{\Phi_0})\cos\eta.\nonumber
\end{eqnarray}

\section{Single qubit manipulation and two-qubit phase gate}

Any unitary transformation can be constructed from a set of basic elementary gates. We
adopt the scheme for single qubit manipulation in Ref.\cite{liu}. However, different
from the flux-qubit in Ref.\cite{liu}, two-qubit gate in our scheme is realized by
adjusting the gate voltage and the externally applied flux.

{\bf Single qubit manipulation can be easily implemented.} In the Hamiltonian
Eq.(\ref{jc}), which has been well studied in quantum optics \cite{walls}, in the large
detuning limit, that is $|\Delta|=|\Omega-\omega|\gg|\lambda|$, the interaction between
qubit and STLR can be neglected  \cite{sun}. In this regime, the single qubit
manipulation may be implemented by changing $n_g$ and $\Phi_e$ rapidly, as if the
superconducting transmission line were absent. It has already been demonstrated in
general in experiments  \cite{nakamura, pashkin, yamamoto}.

{\bf We now study the two-qubit gate}. By controlling gate voltage
and the externally applied flux, one can set a qubit resonating with
the STLR, that is $\Delta=0$. In the resonant case, the evolutions
of the states of qubit and STLR are
\begin{subequations}\label{evolution}
\begin{eqnarray}
&&|g,0\rangle\rightarrow|g,0\rangle,\\\nonumber
&&|g,n+1\rangle\rightarrow\\\label{r1} &&[\cos(\sqrt{n+1}\lambda
t)|g,n+1\rangle-i\sin(\sqrt{n+1}\lambda t)|e,n\rangle],\\\nonumber
&&|e,n\rangle\rightarrow\\\label{r2} &&[\cos(\sqrt{n+1}\lambda
t)|e,n\rangle-i\sin(\sqrt{n+1}\lambda t)|g,n+1\rangle].
\end{eqnarray}
\end{subequations}
Here, in Eq.(\ref{r1}) and (\ref{r2}) we dropped a global phase
factor $e^{-i(n+1)\omega t}$.

\begin{figure}[htbp]

\includegraphics[width=3in]{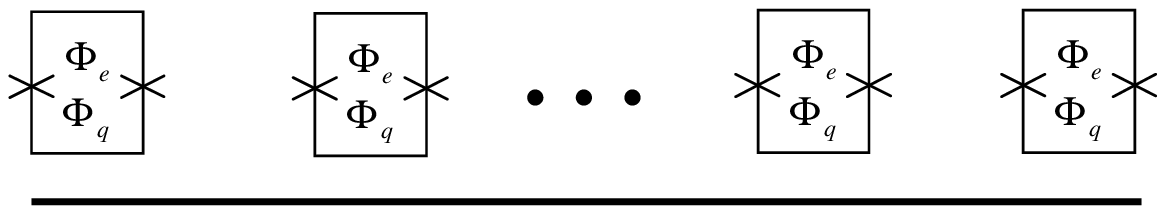}
\caption{The N qubits are placed at the points $x=(2n+1)l/(2k)$
along the STLR, $n$ is an arbitrary integer. $\Phi_e$ and $\Phi_q$
are externally applied flux and flux induced by the transmission
line.}\label{scalable}
\end{figure}

A scalable quantum circuit can be constructed by placing $N$ charge qubits at the
points $x=(2n+1)l/(2k)$ along the STLR acting as the data bus, as shown in
Fig.\ref{scalable}, and a similar setup where the qubits are connected by capacitances
has been given in Ref. \cite{wangyd} where phase transition  has been studied in
detail. The Hamiltonian of the system reads
\begin{equation}
H_s=\hbar\omega(a^{\dagger}a+\frac{1}{2})-\frac{1}{2}\hbar\sum_{k=1}^{N}
\Omega_k\sigma_z^{(k)}+\hbar\sum_{k=1}^{N}
\lambda_k(a^{\dagger}\rho_-^{(k)}+\rho_+^{(k)}a).
\end{equation}

If the detuning between each qubit and STLR is great larger than
their coupling constants, that is $|\lambda_k|/|\Delta_k|\ll1$, then
all N qubits are decoupled from the cavity, so one can do the
single-qubit operations as discussed above.

To implement two-qubit manipulations, the selected two qubits should
be sequentially coupled to the STLR. Suppose we want to implement a
two-qubit manipulation acting on the $i$-th and $j$-th qubits, we
can sequentially setting the the $i$-th and $j$-th resonating with
the STLR. Firstly, we consider how to implement a two-qubit phase
gate. Initially set the STLR in the vacuum state $|0\rangle$, and
let the $i$-th qubit, the $j$-th qubit and then the $i$-th qubit
sequentially resonating with the STLR for time duration $\tau_1$,
$\tau_2$ and $\tau_3$ with $\tau_1=\tau_2=\tau_3=2\pi/\lambda$,
according to formulas (\ref{evolution}), the dynamical evolutions of
the four states $|g_i, g_j, 0\rangle$, $|g_i, e_j, 0\rangle$, $|e_i,
g_j, 0\rangle$, $|e_i, e_j, 0\rangle$ are
\begin{subequations}
\begin{eqnarray}
&|g_i,g_j,0\rangle&\rightarrow
e^{i\theta_1}|g_i,g_j,0\rangle,\\
&|g_i,e_j,0\rangle&\rightarrow
e^{i\theta_2}|g_i,e_j,0\rangle,\\
&|e_i,g_j,0\rangle&\rightarrow
e^{i\theta_3}|e_i,g_j,0\rangle,\\
&|e_i,e_j,0\rangle&\rightarrow e^{i\theta_4}|e_i,e_j,0\rangle
\end{eqnarray}
\end{subequations}
with
\begin{subequations}
\begin{eqnarray}
&\theta_1&=\frac{(\Omega_i+2\Omega_j)\pi}{\lambda},\\
&\theta_2&=-\frac{(-\Omega_i+2\Omega_j+2\omega)\pi}{\lambda},\\
&\theta_3&=-\frac{(\Omega_i-2\Omega_j+4\omega)\pi}{\lambda},\\
&\theta_4&=-\frac{i(\Omega_i+2\Omega_j+6\omega)\pi}{\lambda}.
\end{eqnarray}
\end{subequations}
Here, we assume that eigenfrequencies $\Omega_i$s of qubits are
different when the qubits are not resonant with the STLR. In fact,
the result won't be affected even if the eigenfrequencies
$\Omega_i$s are equal. After the dynamical evolutions with the given
time durations, a two-qubit phase gate is constructed as follows
\begin{eqnarray}
U_{p}= \left(
\begin{array}{cccc}
e^{i\theta_1} & 0 & 0 & 0 \\
0 & e^{-i\theta_2} & 0 & 0 \\
0 & 0 & e^{-i\theta_3} & 0 \\
0 & 0 & 0 & e^{-i\theta_4}
\end{array}
\right)
\end{eqnarray}
Because all of the two-qubit gates are universal  \cite{deutsch}, any two qubit
operation can be obtained by combining the two-qubit phase gate $U_{p}$  with
well-chosen single-qubit operations for the $i$-th and $j$-th qubits. Therefore,
quantum computation can be realized by combination of two-qubit phase gate with
single-qubit operations.

\section{generation of entanglement}

\subsection{two-qubit entanglement}

The entangled state between the $i$-th and $j$-th  qubits can be
easily generated with the assistance of STLR. If two qubits, say the
$i$-th and $j$-th qubits, resonate with the STLR simultaneously, the
Hamiltonian of the system reads
\begin{equation}\label{h2}
H_I=\hbar\omega(a^{\dagger}a+\frac{1}{2})-\frac{1}{2}\hbar\sum_{k=i,
j}\omega\sigma_z^{(k)}+\hbar\sum_{k=i, j}
\lambda_k(a^{\dagger}\rho_-^{(k)}+\rho_+^{(k)}a),
\end{equation}
here, for simplicity, we assumed that the coupling coefficient
$\lambda_{k}$ is the same when qubits resonate with the STLR. The
eigenvalues of Hamiltonian Eq.(\ref{h2}) are
\begin{subequations}
\begin{eqnarray}
&&E_1=\hbar\omega(n-\frac{1}{2})-\hbar\lambda\sqrt{4n-2},\\
&&E_2=E_3=\hbar\omega(n-\frac{1}{2}),\\
&&E_4=\hbar\omega(n-\frac{1}{2})+\hbar\lambda\sqrt{4n-2},\\\nonumber
\end{eqnarray}
\end{subequations}
and the corresponding eigenvectors are
\begin{subequations}\label{eigenvector}
\begin{eqnarray}
\nonumber
|\psi_1\rangle&=&-\sqrt{\frac{n}{4n-2}}|g,g,n\rangle+\frac{1}{2}|g,e,n-1\rangle\\
&&+\frac{1}{2}|e,g,n-1\rangle-\sqrt{\frac{n-1}{4n-2}}|e,e,n-2\rangle,\\
|\psi_2\rangle&=&-\sqrt{\frac{n-1}{2n-1}}|g,g,n\rangle+\sqrt{\frac{n}{2n-1}}|e,e,n-2\rangle,\\
|\psi_3\rangle&=&-\frac{1}{\sqrt2}|g,e,n-1\rangle+\frac{1}{\sqrt2}|e,g,n-1\rangle,\\\nonumber
|\psi_4\rangle&=&\sqrt{\frac{n}{4n-2}}|g,g,n\rangle+\frac{1}{2}|g,e,n-1\rangle\\
&&+\frac{1}{2}|e,g,n-1\rangle+\sqrt{\frac{n-1}{4n-2}}|e,e,n-2\rangle.\\\nonumber
\end{eqnarray}
\end{subequations}
The ground state of Hamiltonian Eq.(\ref{h2}) is $|g,g,0\rangle$ and the corresponding
energy is $E=-\frac{1}{2}\hbar\omega$. From the eigen-states, it is easy to write out
the evolution of the system. For instance if the initial state of the system is
$|g,g,1\rangle$, the state at time $t$ will be
\begin{eqnarray}\nonumber
|\psi(t)\rangle&=&e^{-\frac{i\omega
t}{2}}[\cos(\sqrt2\lambda t)|g,g,1\rangle-\frac{i}{\sqrt2}\sin(\sqrt2\lambda t)|g,e,0\rangle\\
&&-\frac{i}{\sqrt2}\sin(\sqrt2\lambda t)|e,g,0\rangle].\label{bell}
\end{eqnarray}
When $\sqrt{2}\lambda t=\pi/2$, the state of the $i$-th and $j$-th
qubits is a Bell-state becomes a Bell-state
$|\Psi_+\rangle=(|g,e\rangle+|e,g\rangle)/{\sqrt2}$. Usually, the
state of the STLR is in the vacuum state $|0\rangle$, to prepare the
system in state $|g,g,1\rangle$, we first start from state
$|e,g,0\rangle$,  then setting $i$-th qubit resonating with STLR for
a time duration $\tau_1=\pi/(2\lambda)$, and according to Eq.
(\ref{r2}), the state becomes $|g,g,1\rangle$. Then the Bell-state
is obtained by setting $i$-th and $j$-th qubit resonating
simultaneously with the STLR for a period of
$\tau_2=\pi/(2\sqrt2\lambda)$, as shown in Eq. (\ref{bell}).

\subsection{Generation of $W$ states}

In this section, we discuss the generation of $W$ states of arbitrary three qubits.
Using sequential resonance between qubits and STLR, the $W$ states can be generated in
a similar way to the generation of Bell states. We consider the $i$-th, $j$-th and
$k$-th qubits. Suppose the initial states of the three qubits $i$, $j$, $k$ and STLR
are $|g\rangle$, $|g\rangle$, $|e\rangle$ and vacuum state $|0\rangle$, respectively.
Firstly, setting the $k$-th qubit resonating with STLR for a time duration $t_1$,
according to Eq.(\ref{evolution}), the evolution of  state of the $k$-th qubit and STLR
is
\begin{equation}\nonumber
|e\rangle_k|0\rangle\rightarrow\cos(\lambda
t_1)|e\rangle_k|0\rangle-i\sin(\lambda t)|g\rangle_k|1\rangle,
\end{equation}
here, we neglect a global phase factor $e^{-i\omega t_1}$. Secondly,
setting the $i$-th and $j$-th qubit resonating with STLR for a time
duration $t_2$, according to Eq.(\ref{eigenvector}), the state of
the combined system (three qubits and STLR) becomes
\begin{eqnarray}\nonumber
&&e^{\frac{i\omega t_2}{2}}\cos(\lambda
t_1)|g\rangle_i|g\rangle_j|e\rangle_k|0\rangle\\ \nonumber
&&-ie^{-\frac{i\omega t_2}{2}}\sin(\lambda t_1)[\cos(\sqrt2\lambda
t_2)|g\rangle_i|g\rangle_j|g\rangle_k|1\rangle\\ \nonumber
&&-\frac{i}{\sqrt2}\sin(\sqrt2\lambda
t_2)|g\rangle_i|e\rangle_j|g\rangle_k|0\rangle\\
&&-\frac{i}{\sqrt2}\sin(\sqrt2\lambda
t_2)|e\rangle_i|g\rangle_j|g\rangle_k|0\rangle].
\end{eqnarray}
If the resonating time $t_2$ satisfies $\cos(\sqrt2\lambda
t_2)=\pi/2$, that is $t_2=\pi/(2\sqrt2\lambda)$, the state of the
combined system reads
\begin{eqnarray}\nonumber
&&[\cos(\lambda t_1)|g,g,e\rangle\\
&&-\frac{1}{\sqrt2}\sin(\lambda
t_1)e^{-i\frac{\omega\pi}{2\sqrt2\lambda}}(|g,e,g\rangle+|e,g,g\rangle)]|0\rangle,
\end{eqnarray}
where we neglect a global phase factor
$e^{i\omega\pi/(4\sqrt2\lambda)}$ and the subscripts $i$, $j$, $k$.
If the resonating time $t_1$ satisfies $\cos(\lambda t_1)=\sqrt3/3$,
we can get the $W$ state of the three qubits
\begin{equation}
\frac{1}{\sqrt3}[|g,g,e\rangle\mp
e^{-i\frac{\omega\pi}{2\sqrt2\lambda}}(|g,e,g\rangle+|e,g,g\rangle)],
\end{equation}
the signs $\mp$ depends on the value of $\sin(\lambda t_1)$.

\section{discussion and conclusion}

It's notable that the coupling coefficient $\lambda$ is different in
the single-qubit operation and two-qubit gate operation, because
$\lambda$ is the function of $\sin(\pi\Phi_e/\Phi_0)\cos\eta$;
moreover, $\Phi_e$ and $\eta$ are external controlled and different
between single-qubit operation and two-qubit gate operation.
However, the single qubit operation  will not be affected so long as
the condition $|\lambda|/|\Delta|\ll1$ is satisfied during the
single-qubit operation, and $\lambda$ can be tuned to its maximum
value when the two-qubit gate operation begins. In this paper, we
consider the current coupling between the superconducting charge
qubit and the STLR, however, the same result can be achieved by
voltage coupling.

In conclusion, we have presented a scheme for scalable superconducting charge qubits
with the assistance of STLR playing the role of data bus. The coupling between the
selected qubit and the STLR can be turned on and off by just controlling the gate
voltage and externally applied flux of the superconducting qubit. As a result, the
single-qubit manipulation can be performed when the qubit decoupled with the STLR, and
two-qubit gate can be implemented by setting two qubits sequentially resonating with
the STLR. The entanglement state between arbitrary two qubits and $W$ state of
arbitrary three qubits can be generated easily and quickly. The operation time in our
scheme is much shorter than the decoherence time of superconducting charge qubits
according to recent experiments  \cite{nakamura, wallraff}. Different from the flux
qubit gate case where two-qubit gate needs a TDEF, both single- and two- qubit gates
are implemented by changing the gate voltage and the external flux in our scheme, the
same set of devices can be used for both of them, which is appealing to experiment.

\begin{acknowledgments}
We thank J. S. Liu for helpful discussions. This work is supported by the National
Fundamental Research Program Grant No. 2006CB921106, China National Natural Science
Foundation Grant Nos. 10325521, 60433050, 60635040,  the SRFDP program of Education
Ministry of China, No. 20060003048 and the Key grant Project of Chinese Ministry of
Education No.306020.
\end{acknowledgments}

\end{document}